\documentstyle[preprint,aps]{revtex}
\begin{document}
\draft
\title{
Non-collinear magnetism in Al-Mn topologically disordered systems
}

\author{
\bf{A. V. Smirnov}$^{(a)}$\cite{adr}
and
 \bf{A. M. Bratkovsky}$^{(b)}$\cite{corr}
}

\vspace{.2in}
\address{$^{(a)}$Institut f\"ur Festk\"orperphysik, Technische Hochschule,
	D-64289 Darmstadt, Germany
}

\address{$^{(b)}$Oxford University, Department of Materials,
	 Oxford OX1 3PH, England
}

\date{November 21, 1995}
\maketitle

\begin{abstract}
We have performed the first ab-initio calculations of a possible complex
non-collinear magnetic structure in aluminium-rich Al-Mn liquids
within the real-space tight-binding LMTO method.
In our previous work we predicted  the existence of large magnetic
moments in Al-Mn liquids
[A.M. Bratkovsky, A.V. Smirnov, D. N. Manh, and A. Pasturel,
\prb {\bf 52}, 3056 (1995)] which has been very recently
confirmed experimentally. Our present calculations show that
there is a strong tendency for the moments on Mn to have a
non-collinear (random) order retaining their large value of about
3~$\mu_B$. The d-electrons on Mn demonstrate a pronounced
non-rigid band behaviour which cannot be reproduced within
a simple Stoner picture.
The origin of the magnetism in these systems is a
topological disorder which drives the moments formation and
frustrates their directions in the liquid phase.
\end{abstract}

\begin{flushleft}
fluctuations, heavy fermions.
\end{flushleft}

\newpage

The behaviour of magnetic atoms dissolved in simple metals
is an active field of research.
The magnetic properties of Mn dissolved in an Al matrix
are therefore of particular interest since this is a matter of controversy
and extensive experimental work, partly aimed at
 Al-Mn quasicrystals
\cite{hauser,berger,chernikov,lasjaunias,audier}.
It was long been thought that Mn is unlikely to have a moment in an Al matrix
\cite{friedel} but Cooper and Miljak found that a Mn impurity in fcc Al carries
a large moment of $\mu=3.2\pm 0.2$ which is
apparently screened by $sp$ electrons up to very high temperatures,
suggesting a surprisingly high value of the Kondo temperature,
$T_K=600 K$ \cite{cooper}.

As for the disordered systems,
many authors have reported that at low
temperatures ($T<10K$) only a very small fraction (about 1\%) of Mn sites
in icosahedral (i-) and amorphous (a-) phases of Al-Mn and Al-Mn-Pd systems
is really magnetic and that those atoms have a large moment ($>1.5\mu_B$)
\cite{berger,chernikov,lasjaunias,audier}.
In the temperature range $T=10-300K$ the value of the magnetic moment on Mn in
disordered phases of $Al_{100-x}Mn_x$ alloys varies
from 0.7 $\mu_B$ (for $x=16$)  to 2.4 $\mu_B$ ($x=45$)\cite{hauser},
where authors have assumed all Mn atoms to be magnetic.
Some data have suggested the existence of a
Mn magnetic moment in liquid $Al_{100-x}Mn_x$ of about
$\mu_{eff}/\mu_B \sim 2.9,
3.2$ for $x=20$ and 40 \cite{maret}, respectively.

Previous theoretical studies have given contradictory results for the magnetic
behaviour of Mn in an Al matrix.
In Refs.~\cite{niem,zeller,hoshino,bag} the moment on Mn in fcc Al was
found to have values varying in the interval 1.74-3.26~$\mu_B$,
whereas in calculations \cite{henry} and \cite{feng}
Mn was found to be paramagnetic.
Liu {\em et al.} \cite{feng} found
no moment on Mn in MnAl$_{n}$ clusters with $n<54$, however, in
clusters containing more than one manganese atom the moment appeared.

We have recently performed  ab-initio calculations for
liquid Al$_{100-x}$Mn$_x$ ($x$=14, 20, and 40)
to gain more insight into the problem of Mn
magnetism in a disordered Al host\cite{amb1}.
Our real-space  spin-polarized calculations
have shown unambiguously the formation of a large moment of about 3~$\mu_B$
on Mn in these metallic liquids. We have demonstrated
that the reason for the moment
formation lies in a smearing out of the van Hove dip in the density
of states which removes the moment in c-Al$_6$Mn
in accordance with experiment. It means that {\em topological disorder}
is the origin of the moment formation on Mn in an Al matrix.
Our  findings have recently been confirmed
experimentally by Hippert {\em et al.}\cite{audier}
who investigated the series of alloys Al$_{1-x-y}$Pd$_x$Mn$_y$
and found that a localized moment appears on Mn atoms in the liquid state
and disappears in the solid state. The moment they found is
2.76$\pm$0.01 $\mu_B$ from susceptibility measurements and 2.74$\pm$0.1
from the neutron scattering data, in agreement with our
calculations\cite{amb1}. The authors \cite{audier} have also observed
that, firstly, the moment is {\em independent} of the Mn concentration
thus demonstrating a single atom behaviour.
Secondly, the  magnetic susceptibility {\em increases} with temperature.
The authors~\cite{audier} have speculated it could
be if only a fraction of the Mn atoms
in the liquid bears a localized moment. In this case
about 60\% of the Mn atoms can be non-magnetic only if the
rest of them carry moments of more than 5~$\mu_B$, which is unlikely.

In the present study we address the question of the
character of magnetic state in Al-Mn liquids, which could have
orientational disorder owing to the random sign of the
 indirect (RKKY)
interaction between 3$d$ ions in a disordered matrix\cite{coey}.
 As has been indicated in Ref.~\cite{pett}
metallic Mn (and Fe as well)
is close to a {\em disordered local moments} regime because it has a
half-filled $d$-shell and, correspondingly, a large Fermi momentum
and  a short spatial period of the RKKY oscillations.
The antiferromagnetic sign of the Mn-Mn interaction was suggested by
Hauser {et al.}\cite{hauser}
for the case of Al-Mn
amorphous alloys and quasicrystals.
However, in
calculations using the KKR-Green's function method the Mn-Mn
interaction in fcc-Al appeared to be of a {\em ferro}magnetic
sign\cite{hoshino}.
It means that only {\em topological disorder} can produce the random sign
of the RKKY interaction on different Mn sites and, therefore, frustrate
the otherwise ferromagnetic order.

In the present work we have implemented
the method \cite{kubler1} within the ab initio
real-space tight-binding (RSTB) LMTO formalism,
successfully applied before to studies of collinear  magnetism in disordered
Fe-B and Ni-B\cite{amb_feb,amb_feb0,nib},
and Al-Mn systems\cite{amb1}, and we now apply it to self-consistent
calculations of the non-collinear magnetic Al$_{100-x}$Mn$_x$ liquids
with $x$=15, 20, and 40.


In a system with non-collinear magnetic order the electrons
experience an exchange field $V^{xc}_{\sigma\sigma'}({\bf r})$,
which depends on the local orientation $\vec e_R$ $(|\vec e_R|=1)$
of the magnetic moment at each atomic site $R$
and local electron and spin density.

In constructing the {\em ab-initio}  Hamiltonian, $H$,
we have followed the method by O.K.Andersen  \cite{oka84}
and transformed $H$ into a tight-binding form to make use of the
real-space recursion method.
The overlap and Hamiltonian matrices in the tight-binding
LMTO method were
expressed via a {\em two-centre} Hamiltonian $h^{\alpha}$,
which for a non-collinear case takes the following form:
\begin{equation}
h^{\alpha}=c^{\alpha}-E_{\nu}+\sqrt{d^{\alpha}} \left(U S^{\alpha}
U^\dagger\right)
\sqrt{d^{\alpha}},
\label{eq:ha}
\end{equation}
where $S^{\alpha}\equiv S^{\alpha}_{R'L'RL}$ is the spin-independent
matrix of the localized structure
constants, $c^{\alpha}$ and $d^{\alpha}$ are
the matrices of potential parameters,
diagonal in spinor space,  $E_{\nu}$ are the reference energies chosen at
the centres of the respective bands, and
$U$ is the spin-$\frac{1}{2}$
rotation matrix.\cite{kubler1}

In practice, to make use of the recursion method, we have constructed
a nearly-orthonormal  representation starting from the most localized
tight-binding
Hamiltonian $h^{\alpha}$, Eq.(\ref{eq:ha}),
rotated such that we obtain the hamiltonian matrix in the global
coordinate system,

\begin{equation}
H^{\gamma}=U^\dagger \left( E_\nu+h^\alpha
( 1-o^{\alpha}h^{\alpha})^{-1}\right) U =
U^\dagger ( E_{\nu}+h^{\alpha}-h^{\alpha}
o^{\alpha}h^{\alpha}+ \cdots ) U
\label{eq:Hg}
\end{equation}

The local density-of-states matrices $N_{R\sigma,R\sigma'}(E)
=-\frac{1}{\pi}{\rm Im}\langle R\sigma|(E-H^\gamma+{\rm i}0)^{-1}|
R\sigma'\rangle $
have been found by the recursion method
with the hamiltonian $H^{\gamma}$, Eq.(\ref{eq:Hg}).
The orientation of a local spin quantization axis can be found easily
(in the ASA) by diagonalising
the density matrix integrated over the atomic spheres.

In the first instance, we have checked the present method
on the well-studied case of fcc-Fe in order to compare the results with
those calculated by the ASW method \cite{Uhl}
and have found all the results to be consistent with each other.

We have then applied the  RSTB-LMTO method for
60 atom structural models for
liquid Al$_{60}$Mn$_{40}$ and Al$_{80}$Mn$_{20}$, and 56 and 98 atom models
for liquid Al$_{84}$Mn$_{14}$.
The structural models of these Al-Mn systems were constructed by means
of a standard
Monte Carlo method with bond-order potentials\cite{amb1}.
To construct the continued fractions needed for the recursions we have
used up to $\sim 1200$-atom clusters built from our supercells
by applying periodic boundary conditions.

The topological short range order in Al$_{60}$Mn$_{40}$ was found to be
quite different
from that in Al$_{80}$Mn$_{20}$: in the former we have $Z_{\rm MnMn}=3.38$
for the Mn-Mn coordination number, whereas in the latter $Z_{\rm MnMn}$
is just 1.36.
The analysis of bond angles shows some
tendency for Al$_{86}$Mn$_{14}$ and Al$_{80}$Mn$_{20}$ liquids to have
an icosahedral motif, but not for Al$_{60}$Mn$_{40}$\cite{amb1}.

We have found a strong
tendency for magnetic moments on Mn to have large absolute values
and orientational disorder (non-collinear magnetism), so that the net
magnetic moment has a very low average value (Table I).
It is important to note that for c-Al$_6$Mn
our non-collinear calculations yielded a non-magnetic state, in accordance
with our previous discussion of the role of the van Hove singularity
at the Fermi level
in the density of states of this system\cite{amb1}.
 All average values in the present calculations
are close to our
previous results \cite{amb1} based on large structural models and averaged
self-consistent potential parameters (Table I).
Moreover,  in the collinear case we have found
no meaningful changes in the distribution of the local magnetic moments,
although
in \cite{amb1} the values of the Mn magnetic moment are somewhat larger.
For Al$_{84}$Mn$_{14}$ system the averages are in good agreement
for small (N=56) and large (N=98) calculated cells in spite of rather few
statistics for Mn in the former calculation.

The analysis of the densities of states projected onto the local
magnetization axes reveals that
the total electronic density of states (DOS) has a sharp peak
for majority spins
in all liquid Al-Mn alloys at about $-2.5$~eV below the Fermi level (Fig.~1),
and a peak in the unoccupied minority spin band at about +1~eV.
The local projected DOSs are similar to
those calculated in our previous work\cite{amb1}.
The difference between collinear and non-collinear DOS grows with increasing
Mn concentration (Fig.~1). We note that the shape of the
majority/minority DOS reflects a strongly non-rigid band behaviour (Fig.~1)
so that the rigid
band Stoner model is hardly applicable to Al-Mn systems.

We have found that the non-collinear state is
{\em more stable} than the collinear one, being
lower in energy by about 0.025 Ry (Table I).
The average value of the Mn moment in our calculation is almost
independent of the manganese concentration in correspondence with the
experiment \cite{audier}.

In our calculations the distribution of
the absolute values of the Mn moment is asymmetric in Al$_{86}$Mn$_{14}$
and Al$_{80}$Mn$_{20}$, and it is biased towards higher values,
whereas the moment distribution in Al$_{60}$Mn$_{40}$ is symmetric (Fig.~2).
To gain more insight into the spatial distribution of the moments on Mn
we have analyzed  the average cosine of the angle between Mn moments,
$\cos(\theta_{ij})$,
as a function of the distance $R_{ij}$ between them (Fig.~3),
in conjunction with the Mn-Mn partial radial distribution function.
For Al$_{86}$Mn$_{14}$ liquid the nearest neighbours are likely to be
subject to a ferromagnetic exchange
interaction, whereas other studied systems display a definite
antiferromagnetic sign of the interaction between nearest Mn atoms
which changes quickly into ferromagnetic with increasing separation.
The analysis of  $<{\vec e}_i{\vec e}_j>_{\rm Mn}$
for all Mn-Mn neighbours with separations $R_{ij}$
demonstrates a preference for
antiferromagnetic alignment of distant ($4.7 \AA < R_{ij} < 6.1 \AA$)
Mn moments (Table I).
The behaviour of Al-Mn systems is quite different compared to Mn in fcc-Al
where the exchange has a ferromagnetic sign up to the 3rd neighbours
\cite{hoshino}:
topological disorder produces RKKY exchange of
random signs and, therefore, results in the random
directional order of moments on the manganese atoms.


In conclusion, present calculations confirm our earlier
prediction\cite{amb1} that topological disorder is the main
driving force for the formation of  a large ($\mu_{eff} \sim 2.8 \mu_B$)
 magnetic moment on Mn in Al-Mn liquids.
This value is
close to the the single-impurity limit \cite{cooper},
and is not sensitive to interaction with other Mn atoms in the alloy,
as found recently in experiment \cite{audier}.
Our results do not confirm the view that only small fraction of  Mn sites
in disordered system  carry a moment due to a strong local
environment effect, with others being
non-magnetic:  we have found that {\em all} Mn sites
are magnetic in the disordered systems we studied (Table I).

The observed rise in magnetic susceptibility $\chi$
above the melting point in Al-Pd-Mn systems \cite{audier} is quite the
opposite to what is expected from the usual spin-fluctuation
theories where $\chi$ is Curie-like and, therefore, decreases with
temperature\cite{moriya}.
This rise may be
 a fingerprint of Kondo
un\-screening with increasing temperature, but could also be a result
of a variation of the moments distribution (Fig.~2) with temperature
and local environment effects, facts which should be
analysed further.

We predict that Al-Mn liquids have a random magnetic order with predominance
of ferromagnetic interactions for nearest Mn neighbours, and that
non-collinearity
is triggered by random RKKY interaction between solute atoms of Mn in
a disordered Al matrix.


The authors are indebted to V. Heine, J.K\"ubler,
D. Nguyen Manh, A. Pasturel, D. Pettifor,
L.Sandratskii, and M.Uhl for helpful discussions,
AVS has been supported by the Alexander von Humboldt Foundation.




\begin{table}
\caption{
The results for collinear (C) and non-collinear (NC) spin configurations.
 $\mu_{min}\div\mu_{max}$
is the interval spanned by the values of Mn moments;
$<\mu_{\rm Mn}>$ and $<\mu_{\rm Al}>$ are the averages of the
moment values on Mn and Al respectively.
$\mu$ is the value of the average moment per atom.
$<{\vec e}_i{\vec e}_j>_{\rm Mn}$ is
the average cosine of the angle between moments on two neighbouring
Mn atoms. $E_{fm}-E_{nc}$
is the energy difference between ferromagnetic and non-collinear
configurations. All the moments are in units of $\mu_B$.
For collinear calculations the directions of Al moments are opposite to
those of Mn moments.
}

\vspace*{1.cm}
\begin{center}
\begin{tabular}{ccccc}
&&Al$_{60}$Mn$_{40}$ & Al$_{80}$Mn$_{20}$ & Al$_{86}$Mn$_{14}$  \\
\tableline
&$\mu_{min}\div\mu_{max}$&1.96 $\div$ 3.08&1.28 $\div$ 3.39&1.62 $\div$ 3.49\\
&$<\mu_{\rm Mn}>$&2.68&2.72&2.84\\
C&$<\mu_{\rm Al}>$&0.096&0.048&0.039\\
&$\mu$&1.01&0.51&0.37\\
\tableline
&$\mu_{\rm Mn}$\tablenotemark[1] & 2.87& 3.17& 3.29 \\
\tableline
&$\mu_{min}\div\mu_{max}$&2.05 $\div$ 3.43&1.42 $\div$ 3.54&1.86 $\div$ 3.39\\
&$<\mu_{\rm Mn}>$&2.74&2.89&2.88 \\
&$<\mu_{\rm Al}>$&0.062&0.036&0.029\\
NC&$\mu$&0.27&0.23&0.08 \\
&$<{\bf e}_i{\bf e}_j>_{\rm Mn}$&&&  \\
&$R_{ij}<4.7\AA$&0.22&0.42&0.13  \\
&$4.7\AA<R_{ij}<6.1\AA$&-0.09&-0.21&-0.25  \\
\tableline
&$E_{fm}-E_{nc}$ (mRy)&25&24&26 \\

\end{tabular}
\tablenotetext[1]{The result of collinear calculation\cite{amb1}}.
\label{t:tab1}
\end{center}

\end{table}





\begin{figure}
\caption{The spin-polarized electronic density of states for Al-Mn
liquids: (a) Al$_{84}$Mn$_{14}$, (b) Al$_{80}$Mn$_{20}$, and (c)
Al$_{60}$Mn$_{40}$.
Solid line: density of states for a global quantization axis, non-collinear
configuration;
dot-dashed line: the same for a local quantization axis;
dotted line: the density of states per spin for the collinear case.
}
\end{figure}

\begin{figure}   
\caption{The diagram  of the Mn moments distribution
in Al-Mn liquids. $\theta$ is the angle between magnetic moment and
magnetization axis. Vertical line marks the centre of gravity of the
distribution. Note the asymmetry of moment distribution in
Al$_{86}$Mn$_{14}$ and Al$_{80}$Mn$_{20}$.
}

\end{figure}

\begin{figure}     
\caption{The cosine of the angle between two Mn moments (solid line)
and partial Mn-Mn radial distribution functions (dashed line, right axis).
}

\end{figure}

\end{document}